\documentclass{article}
\usepackage{graphicx} % Required for inserting images
\usepackage{lipsum}
\usepackage{amsmath}
\usepackage{makecell}
\usepackage{cite}
\usepackage{hyperref}
\usepackage{xcolor}
\usepackage{multirow}
\usepackage{ulem}
\usepackage[letterpaper,top=1in,bottom=1in,left=1in,right=1in]{geometry}
\usepackage{comment}
\usepackage{booktabs}

\usepackage{lineno}
\usepackage{amsmath}
\usepackage{etoolbox} %% <- for \cspreto, \csappto
\newcommand*\linenomathpatch[1]{%
  \cspreto{#1}{\linenomath}%
  \cspreto{#1*}{\linenomath}%
  \csappto{end#1}{\endlinenomath}%
  \csappto{end#1*}{\endlinenomath}%
}
\newcommand\freefootnote[1]{%
  \let\thefootnote\relax%
  \footnotetext{#1}%
  \let\thefootnote\svthefootnote%
}

\linenomathpatch{equation}
\linenomathpatch{gather}
\linenomathpatch{multline}
\linenomathpatch{align}
\linenomathpatch{alignat}
\linenomathpatch{flalign}
\begin {document}

% --------------------------------------------------------------------
% --------------------------------------------------------------------

\begin{center}
%\vspace*{30pt}
{\huge \bf 
Nuclear Dependence of Beam Normal Single Spin Asymmetry in Elastic Scattering from Nuclei}

\vspace*{10pt}
{\LARGE (A New Proposal Submitted to PAC 52)}

\vspace*{10pt}
\today

\vspace*{20pt}
Alexandre Camsonne, Mark Dalton, Ciprian Gal$^{*\dagger}$, Dave Gaskell, Chandan Ghosh$^{*}$, \\Douglas Higinbotham, Robert Michaels, Sanghwa Park$^{*}$, Anil Panta\\
{\it Jefferson Lab, Newport News, VA, USA}
\vspace*{10pt}

Dustin McNulty\\
{\it Idaho State University, Pocatello, ID, USA}
\vspace*{10pt}

Charles J. Horowitz, Caryn Palatchi\\
{\it Indiana University, Bloomington, IN, USA}
\vspace*{10pt}

Rakitha Beminiwattha, Shashini Chandrasena, Neven Simicevic, \\Steven P. Wells, Lasitha Weliyanga\\
{\it Louisiana Tech University, Ruston, LA, USA}
\vspace*{10pt}

Juliette Mammei\\
{\it University of Manitoba, Winnipeg, MB, Canada}
\vspace*{10pt}

Krishna Kumar\\
{\it University of Massachusetts, Amherst, MA, USA}
\vspace*{10pt}

Tyler Kutz\\
{\it Massachusetts Institute of Technology, Cambridge, MA, USA}
\vspace*{10pt}

Paul King, Arindam Sen\\
{\it Ohio University, Athens, OH, USA}
\vspace*{10pt}

Abhay Deshpande\\
{\it Stony Brook University, Stony Brook, NY, USA}
\vspace*{10pt}

Kent Paschke\\
{\it University of Virginia, Charlottesville, VA, USA}
\vspace*{10pt}

Devi Adhikari, Mark Pitt\\
{\it Virginia Tech, Blacksburg, VA, USA}
\vspace*{10pt}

David Armstrong\\
{\it College of William and Mary, Williamsburg, VA, USA}
\vspace*{10pt}

\freefootnote{*: Cospokespeople, $\dagger$: Contact (ciprian@jlab.org)}
\end{center}
\clearpage

% --------------------------------------------------------------------
% --------------------------------------------------------------------
%\newpage

%\null\vfill
%\hrule

%\linenumbers
%% Since we have executive summary, I think we can skip abstract
%\newpage
%\begin{abstract}
%\lipsum[1-1]
%\end{abstract}

\newpage
\tableofcontents

\newpage
\section{Executive Summary}
%\subsection{Physics goals}
%The goal of this proposal is to resolve the PREX $A_n$ puzzle. The beam normal single spin asymmetry in elastic scattering of nuclei can be described well up to $Z=20$ by the theory of two-photon-exchange (TPE). However, the $A_n$ for lead-208 is consistent with zero in spite of predictions from TPE predicting a large few ppm asymmetry. Recent theoretical proposals have been made that indicate a $Z^2$ scaling of this asymmetry suppression. 
The goal of this proposal is to resolve the so-called PREX $A_{n}$ puzzle where the measured asymmetry for $^{208}$Pb shows a striking disagreement from the theoretical prediction and from the results for $Z \leq$ 20 nuclei. The beam normal single spin asymmetry $A_{n}$ in elastic electron scattering off nuclei has been measured for several target nuclei mostly from light to intermediate mass. The asymmetry arises from two 
%or more %% not sure about this CiG
photon exchange (TPE) as a single photon exchange contribution would vanish under time-reversal symmetry. The PREX $^{208}$Pb results indicate that there are missing contributions that are not accounted for in the existing theoretical models.

%\subsection{Proposed experiment}
We propose to measure the asymmetry using targets with a broad range for $Z$ (from $Z=6$ to $Z=90$) at $Q^{2}$ = 0.0092 GeV$^{2}$. We request 8.6
days of beam time in Hall C using the SHMS spectrometer. The measurement will provide new data for intermediate to heavy nuclei, where no data exists for $Z\geq 20$ in the kinematics of previous high-energy experiments. The TPE predictions for all the nuclei in the proposal sit at approximately 6~ppm. The precision of the collected data will reach 0.5~ppm for each nucleus.

%\subsection{Technical considerations}
This experiment leverages the procedures and techniques developed at the lab for the PREX-2, CREX, and Qweak experiments. The systematic requirements are less stringent compared to what was obtained for the equivalent $A_n$ measurements during PREX-2 and CREX. We estimate that the total systematic budget should be smaller than 0.2~ppm, which is less than half of the statistical uncertainty for each target.

\newpage
\section{Introduction}

\subsection{Physics overview}
%General overview of two-photon exchange and asymmetry. (Cip)\\
Beam normal single spin asymmetries (also known as vector analyzing powers) for relativistic scattering of electrons from nuclei have been measured at many facilities around the world and are considered well-understood for light and intermediate nuclei. The asymmetry arises as an azimuthal modulation of the scattering amplitudes when the electron is polarized transversely to the beam momentum direction. The effect is due to the interaction between the electron's dipole moment and the nucleus' magnetic field in the electron's rest frame. The asymmetry can be defined as:
\begin{equation}
    A(\phi) = A_n (\theta,E_{\rm beam}) P_n \cos(\phi),
\end{equation}
where $\phi$ and $\theta$ are the azimuthal and scattering angles respectively, and $E_{\rm beam}$ is the energy of the electron beam. For an electron beam polarization $\vec{P}$ with $\hat{n}$ as the unit vector normal to the scattering plane given by ${(\vec{k}\times\vec{k'})}/(|\vec{k}\times\vec{k'}|)$ with $\vec{k} (\vec{k'})$ the momentum of the incoming (outgoing) electron, then $P_n \cos(\phi) =  \vec{P}\cdot\hat{n} $. At ultra-relativistic electron beam energies, the internal structure of the target nucleus can play an important role in this asymmetry.

This observable is a measure of higher order effects. Since scattering will be invariant under time reversal, these asymmetries should vanish in first-order Born approximation.
The dominant contribution then is the interference between the imaginary part of the two-photon exchange amplitude(TPE - see Figure~\ref{fig:prex_radCorr} (a) ) and the one-photon exchange amplitude~\cite{DeRujula:1972te}:
\begin{equation}
    A_n = \frac{2A_{Born}\mathcal{I}(A^*_{2\gamma})}{\left| A_{Born} \right|^2},
\end{equation}
where $A_{Born}$ is the purely real one-photon exchange amplitude, and $\mathcal{I}(A^*_{2\gamma})$ is the imaginary (absorptive) part of the two-photon exchange amplitude $A_{2\gamma}$.
%\footnote{This is due to the time-reversal invariance of the electromagnetic interaction.}

For experiments with small scattering angles, a convenient approach to calculate $A_n$ involves using the optical theorem to relate the virtual photoabsorption cross section to the doubly-virtual Compton scattering amplitude~\cite{Afanasev:2004pu,Gorchtein:2004ac,Gorchtein:2005yz,Gorchtein:2005za}. This approach intrinsically includes all excited intermediate states.
More sophisticated calculations such as the recent one performed by Koshchii~{\it et al.}~\cite{Koshchii:2021mqq} have solved the Dirac equation numerically and the contribution of inelastic intermediate states was included in the form of an optical potential with an absorptive component. In addition, the Compton slope parameter was made $A$-dependent based on experimental data. However, even this calculation still is not able to describe the measured data (see Figure~\ref{fig:prex_at}), although we do note that at 2~GeV the corrections applied to the calculation do seem to have an effect.

\subsection{Previous measurements}
%Briefly summarize results from Mainz, PREX, CREX ...
For this proposal we will limit our overview to nuclear scattering, however we note that several measurements have been performed to determine $A_n$ for the proton (see~\cite{Wells:2000rx,Armstrong:2007vm,Androic:2011rh,Maas:2004pd,Gou:2020viq,Rios:2017vsw,Androic:2020rkw}). 

Measurements of $A_{n}$ on nuclei have been performed at Mainz and Jefferson Lab, for beam energies from 0.570 to 3~GeV and scattering angles from about 5 to 26 degrees. The existing $A_{n}$ results are summarized in Table~\ref{tab:prevResults}.
%The HAPPEX and PREX collaborations expanded the study of $A_n$ to nuclei, with measurements on $^{4}\textrm{He}$, $^{12}\textrm{C}$, and $^{208}\textrm{Pb}$ at forward angle, 6\degree, and energies 1~-~3~GeV~\cite{Abrahamyan:2012cg}.  
%The A1 collaboration reported results on $^{12}\textrm{C}$ with a beam energy of 570 MeV and larger angles (15\degree--26\degree) for a variety of $Q^2$~\cite{Esser:2018vdp}, and from $^{28}$Si and $^{90}$Zr with the same energy and similar angles~\cite{Esser:2020vjb}.  The Q$_{\rm weak}$ collaboration reported results on $^{12}\textrm{C}$ and $^{27}\textrm{Al}$ at an angle of 7.7$^\circ$~\cite{PhysRevC.104.014606}.
\begin{table}[!ht]
    \centering
    \begin{tabular}{c|l|c|c|c|c}
    \hline
     Experiment & Target & E$_{beam}$ (GeV) & Angle (deg) & Q$^{2}$ (GeV$^2$) & $A_{n}$ (ppm)\\
     \hline
     PREX-II & $^{12}$C   & 0.95  & 4.87 & 0.0066 & -6.3 $\pm$ 0.4 \\
     \& CREX & $^{40}$Ca  & 0.95 & 4.81 & 0.0065 & -6.1 $\pm$ 0.3 \\
      \cite{PREX:2021uwt} & $^{208}$Pb & 0.95 & 4.69 & 0.0062 & 0.4 $\pm$ 0.2 \\
      & $^{12}$C   & 2.18 & 4.77 & 0.033 & -9.7 $\pm$ 1.1 \\
      & $^{40}$Ca  & 2.18 & 4.55 & 0.030 & -10.0 $\pm$ 1.1 \\
      & $^{48}$Ca  & 2.18 & 4.53 & 0.030 & -9.4 $\pm$ 1.1 \\
      & $^{208}$Pb & 2.18 & 4.60 & 0.031 & 0.6 $\pm$ 3.2 \\
      \hline
     HAPPEX    & $^{4}$He   & 2.750 & 6 & 0.0773 & -13.97 $\pm$ 1.45\\
     \& PREX-I & $^{12}$C   & 1.063 & 6 & 0.00984 & -6.49 $\pm$ 0.38\\
       \cite{Abrahamyan:2012cg}        & $^{208}$Pb & 1.063 & 6 & 0.00881 & 0.28 $\pm$ 0.25\\
    \hline
    Mainz & $^{12}$C  & 0.570 & 15.10 & 0.023 & -15.984 $\pm$ 1.252 \\
     \cite{Esser:2018vdp,Esser:2020vjb}     & $^{12}$C  & 0.570 & 23.50 & 0.039 & -23.877 $\pm$ 1.225 \\
          & $^{28}$Si & 0.570 & 23.51 & 0.038 & -23.302 $\pm$ 1.470 \\
          & $^{28}$Si & 0.570 & 19.40 & 0.036 & -21.807 $\pm$ 1.480 \\
          & $^{90}$Zr & 0.570 & 23.51 & 0.042 & -17.033 $\pm$ 3.848 \\
          & $^{90}$Zr & 0.570 & 20.67 & 0.042 & -16.787 $\pm$ 5.688 \\
    \hline
     Qweak    & $^{12}$C   & 1.158 & 7.7 & 0.02528 & -10.68 $\pm$ 1.07\\
    \cite{PhysRevC.104.014606}       & $^{27}$Al   & 1.158 & 7.7 & 0.02372 & -12.16 $\pm$ 0.85\\
    \hline
    \end{tabular}
    \caption{$A_{n}$ measurements for A $>$ 1 nuclei.}
    \label{tab:prevResults}
\end{table}

\iffalse
\begin{itemize}
    \item Mainz: $^{12}\textrm{C}$~\cite{Esser:2018vdp}, $^{28}$Si and $^{90}$Zr~\cite{Esser:2020vjb}
    \item HAPPEX: $^{4}\textrm{He}$, $^{12}\textrm{C}$,$^{208}\textrm{Pb}$ ~\cite{Abrahamyan:2012cg}
    \item Qweak: $^{12}\textrm{C}$, $^{27}\textrm{Al}$~\cite{PhysRevC.104.014606}
    \item PREX, CREX: $^{12}\textrm{C}$, $^{40}\textrm{Ca}$, $^{48}\textrm{Ca}$, $^{208}\textrm{Pb}$~\cite{PREX:2021uwt}
\end{itemize}
\fi

\begin{figure}[!ht]
    \centering
    \includegraphics[width=0.8\textwidth]{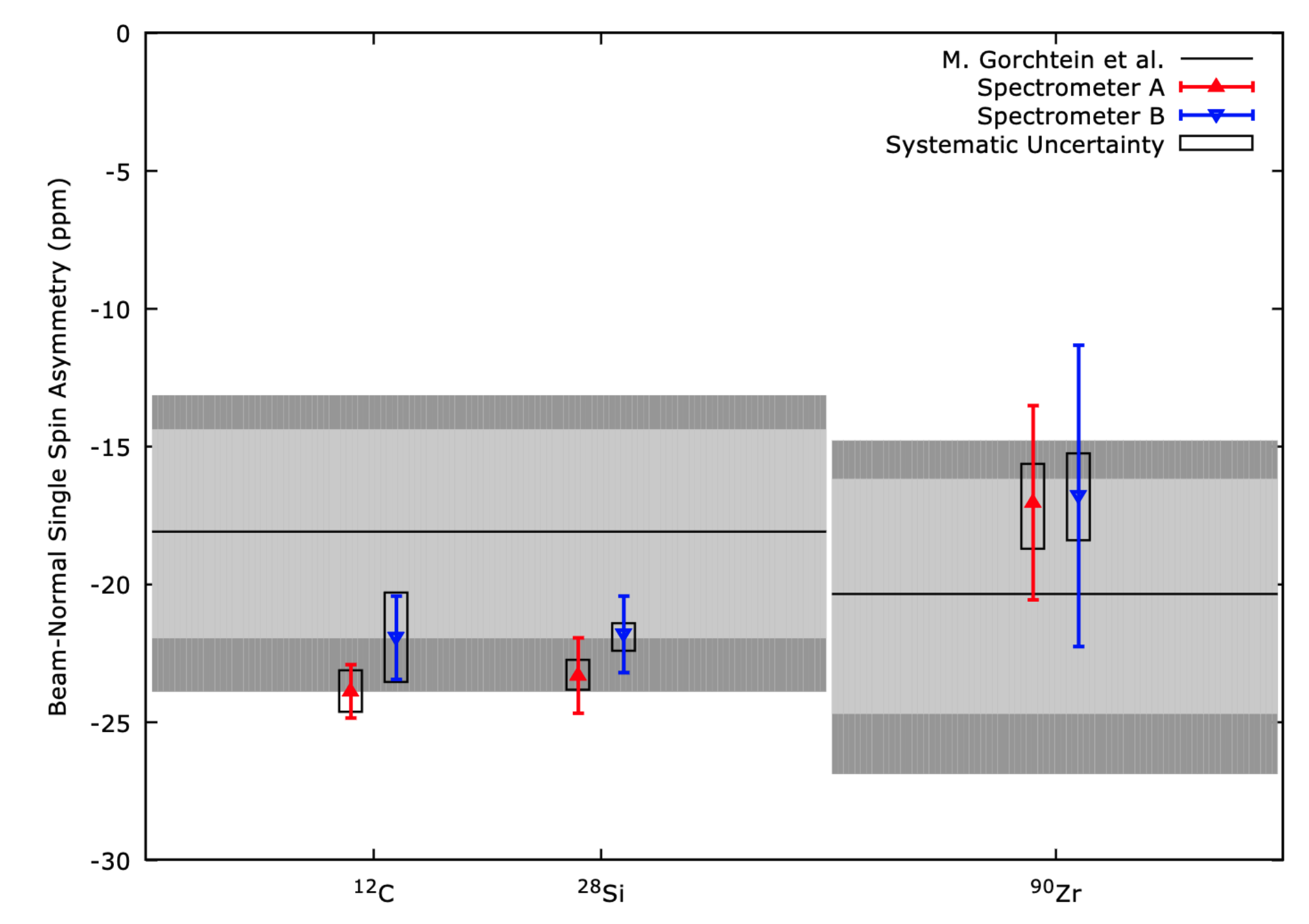}
    \caption{$A_{n}$ measurements from Mainz (figure 3 of~\cite{Esser:2020vjb}). Shaded bands indicate the theoretical uncertainty on the $A_n$ calculation.}
    \label{fig:mainzZr}
\end{figure}

The Mainz measurements were done for a variety of momentum transfers, but with larger angles (15--24 degrees) making the use of the optical theorem approach for comparisons with TPE calculations more difficult. As can be seen in Figure~\ref{fig:mainzZr} the large uncertainties (both experimental and theoretical) make it difficult to make a clear differentiation between the low-Z and intermediate-Z nuclei results.

\begin{figure}[!ht]
    \centering
    \includegraphics[width=0.7\textwidth]{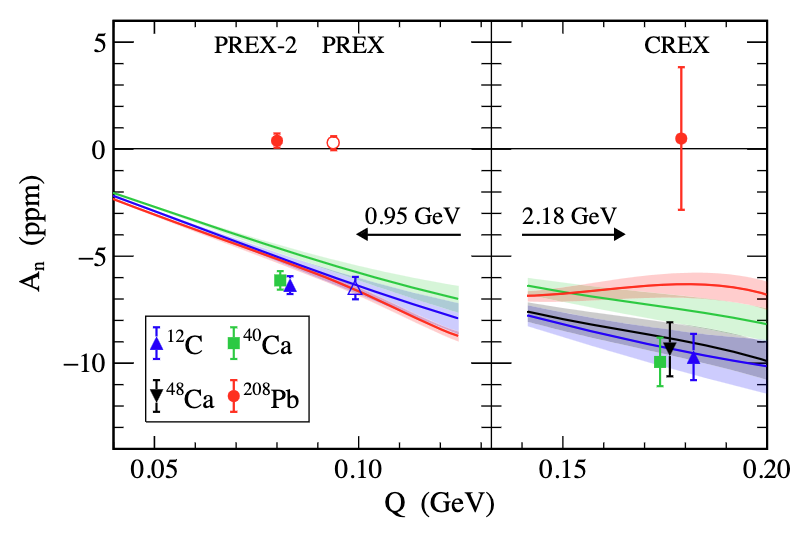}
    \caption{$A_{n}$ measurements from PREX, PREX-2 and CREX~\cite{PREXCREXAT:2022}. The solid lines show theoretical calculations from~\cite{Koshchii:2021mqq} at 0.95\;GeV and 2.18\;GeV, while the shaded region indicates the theoretical uncertainties associated with each calculation.}
    \label{fig:prex_at}
\end{figure}

In contrast, the JLab results provide a relatively easier theoretical interpretation as can be seen in the uncertainty bands from Figure~\ref{fig:prex_at}.
The measured $A_n$ by the PREX and CREX collaborations are at forward scattering angles around 5 degree and 1--2~GeV beam energies. As can be seen, the low-Z nuclei (C and Ca) are in good agreement with the TPE calculations, however, the Pb results provide a theoretical mystery. The multiple $Q^2$ measurements all favor a slightly positive $A_n$ in stark contrast to the TPE calculations (including the most sophisticated ones performed by Koshchii~{\it et al.}~\cite{Koshchii:2021mqq}). This has been dubbed the "PREX puzzle". The resolution of this puzzle is the main thrust of this proposal.

\subsection{Verifiable hypotheses}
\label{sec:hypotheses}
We discuss two hypotheses below that may explain the observed nuclear dependence of $A_{n}$. Both of the hypotheses assume a $Z^2$ scaling of the asymmetry ($A_n = A_0(Q^2)(1 - C \cdot (Z\alpha)^2)$) and would be tested by the data collected with this experiment. The full set of measurements we propose here could show the turn-on of a process that competes with the two-photon exchange contribution to the transverse asymmetry leading to a complete cancellation by the time it reaches lead ($Z=82$).

Of crucial importance is that there are no theories that account for a change in the $A_n$ at 1~GeV. Table~\ref{tab:prexData} shows the results obtained at this energy with PREX-2. The low-Z results (carbon and calcium) are consistent with each other, however, they are in stark contrast to the lead result presenting a difference of 21$\sigma$.
\begin{table}[ht]
    \centering
    \begin{tabular}{ccccc}
      \toprule 
          E$_{\rm beam}$ &  \multirow{2}{*}{Target} & \multirow{2}{*}{$\rm A_n$ (ppm)} & \multirow{2}{*}{$A_\text{avg}^{Z\leq20}$ (ppm)} & \multirow{2}{*}{$\frac{\rm A_n - A_{avg}^{Z\leq20}}{\rm uncert}$}  \\
           ${\rm (GeV)}$ & & & & \\
     \midrule
          0.95 & $^{12}$C   & \multirow{2}{*}{$\left.\begin{array}{c}
          \ -6.3\pm0.4 \\
          \ -6.1\pm0.3 
          \end{array}\right\rbrace$}   &\multirow{2}{*}{$-6.2\pm0.2$} & \\
          0.95 & $^{40}$Ca  &  & & \\
     \midrule
          0.95 & $^{208}$Pb & $\ 0.4\pm0.2$  & & 21~$\sigma$ \\
    \bottomrule
    \end{tabular}
    \caption{Reproduction of table III in \cite{PREXCREXAT:2022} showing $A_n$ results for the three nuclei measured at 0.95~GeV along with the corresponding total uncertainties (statistical and systematic uncertainties combined in quadrature).}
    \label{tab:prexData} 
\end{table}

\subsubsection{Radiative correction}
\begin{figure}[!ht]
    \centering
    \includegraphics[width=0.75\textwidth]{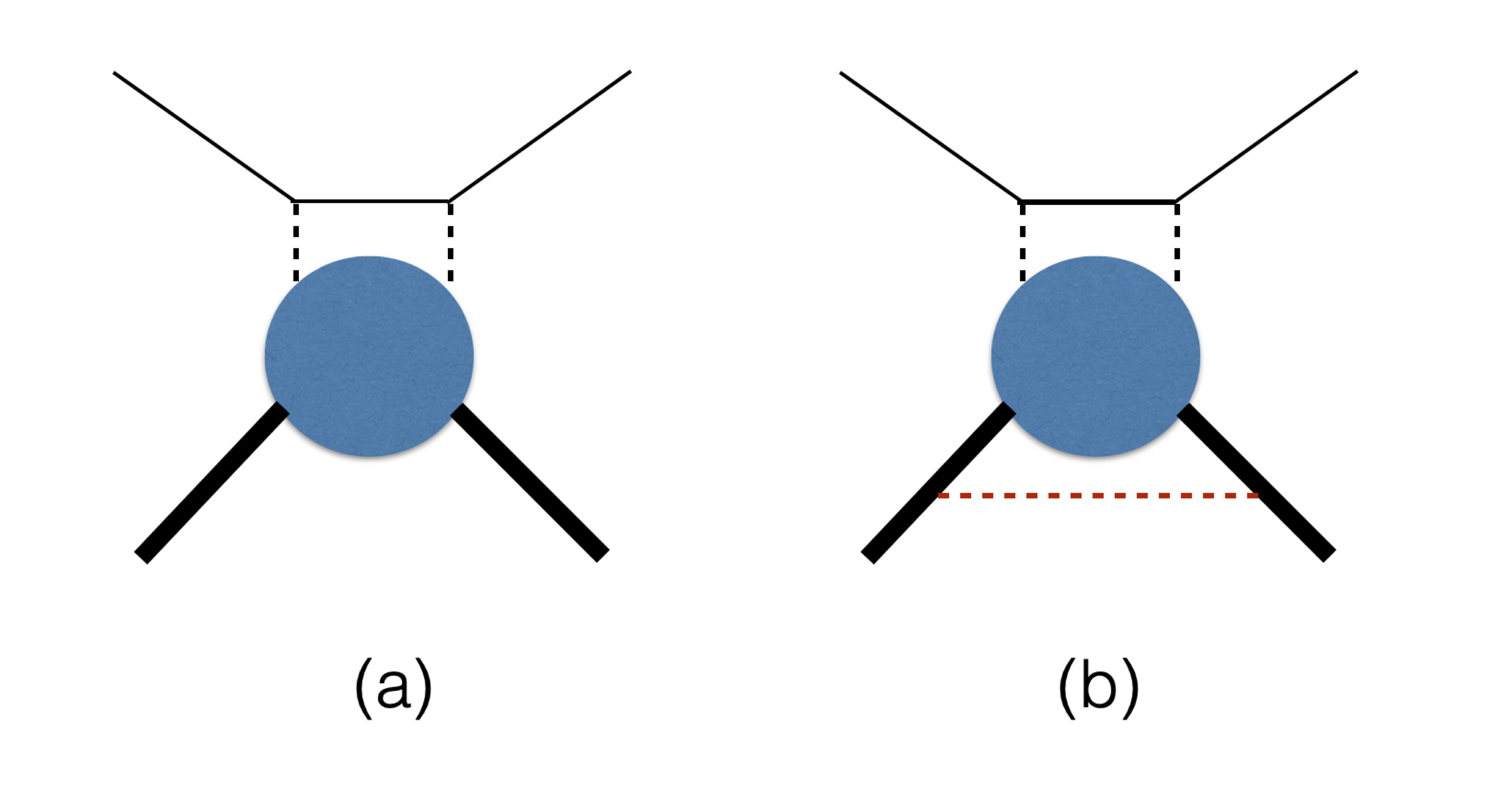}
    \caption{Illustrations of (a) two-photon exchange process and (b) the process with radiative corrections that could modify the observed asymmetry in Pb but leave low Z measurements untouched~\cite{PREXCREXAT:2022}.}
    \label{fig:prex_radCorr}
\end{figure}

% discuss Z^2 scaling from the PREX-2/CREX AT paper
As detailed in \cite{PREXCREXAT:2022} there are no current calculations that can accommodate the suppression of the Pb results but leave the low-Z measurements untouched. An explanation proposed by the authors of that paper would be a correction such as the one presented in panel (b) of Figure~\ref{fig:prex_radCorr}. The correction is the red-dashed line and would provide a scaling of Z$^2$ due to the interaction with both the incoming and outgoing legs of the nucleus. As pointed out in the paper a correction of this type 
would be consistent with the PREX-1, PREX-2, CREX data as well as the Zr data from Mainz (although the large uncertainty on the Zr data does not provide conclusive evidence).

\subsubsection{Beyond the Standard Model physics}
% discuss BSM connection
An alternative way for the two-photon exchange contribution for lead (and to a lesser degree zirconium) to be suppressed would be due to a competing physics process with an opposite sign effect. 

% review https://arxiv.org/pdf/2301.02304.pdf
In their recent work~\cite{Boughezal:2023ooo} Boughezal and collaborators detail exactly such a process. They have identified two dipole operators in the SMEFT framework that provide a non-vanishing transverse asymmetry through a chirality flip\footnote{For all the operators they have considered see eq 17 in their paper and the discussion around it.} (see equation~\ref{eq:dipoleOp}). These dipole contributions arise in transversely polarized electron deep inelastic scattering from unpolarized protons. In equations~\ref{eq:dipoleOp} $W^I$ and $B$ are the SM field strength tensors for the SU(2) and U(1) gauge groups, and $\tau^I$ are the Pauli matrices.

\begin{align}
\begin{split}
    \mathcal{O}_{\rm eW} &= \left( \bar{l} \sigma^{\mu \nu} e \right) \tau^I \phi W^I_{\mu \nu}  \\
    \mathcal{O}_{\rm eB} &= \left( \bar{l} \sigma^{\mu \nu} e \right) \phi B_{\mu \nu} 
%    \mathcal{O}_{\rm uW} &= \left( \bar{q} \sigma^{\mu \nu} u \right)  \tau^I \phi W^I_{\mu \nu}  \\
%    \mathcal{O}_{\rm uB} &= \left( \bar{q} \sigma^{\mu \nu} u \right) \phi B_{\mu \nu} \\
%    \mathcal{O}_{\rm dW} &= \left( \bar{q} \sigma^{\mu \nu} d \right)  \tau^I \phi W^I_{\mu \nu}  \\
%    \mathcal{O}_{\rm dB} &= \left( \bar{q} \sigma^{\mu \nu} d \right) \phi B_{\mu \nu}     
\end{split}
\label{eq:dipoleOp}
\end{align}

\begin{figure}[!ht]
    \centering
    \includegraphics[width=0.75\textwidth]{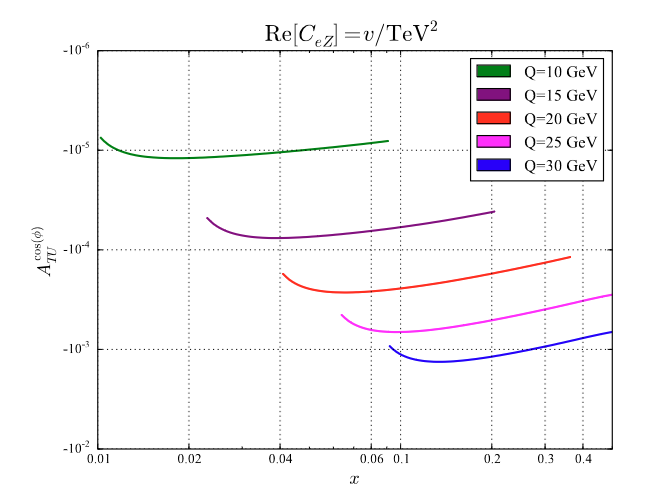}
    \caption{Right panel of figure 2 in~\cite{Boughezal:2023ooo} showing the contribution of dipole operators to the transverse asymmetry as a function of Q$^2$ (different colored lines) and Bjorken-x.}
    \label{fig:smeft}
\end{figure}

The resulting transverse asymmetry for a $Q^2$ of 900~GeV$^2$ from these operators would be on the level of 1000~ppm for new particles with masses in the range of TeV energies as can be seen in Figure~\ref{fig:smeft}. The corresponding two-photon exchange asymmetry in this case is on the level of 0.1~ppm. The choice of TeV energies for the new particle masses shows that probing such a process would lead to significant constraints on BSM models beyond what is currently envisioned to be reached in the next decade. The impact of these operators on the transverse asymmetry scales with Q$^2$~\cite{petrieloPrivateComm}.

% estimate low q2 effect and how it compares to figure 1
Scaling the contribution from these operators to the low energy regime of PREX and the proposed experiment here results in a reduction of approximately 5 orders of magnitude. This would bring the asymmetry contribution to the level of 10~ppb. However, at low momentum transfers we can take advantage of a Z$^2$ cross-section enhancement. We follow the work presented in~\cite{Liu:2017htz,Davoudiasl:2021mjy,Davoudiasl:2023pkq} which are focused on axion production in eA collisions. In particular they treat the nucleus as a scalar, leading to a photon-nucleus vertex that is given by
\begin{equation}
iV^{\mu}(q^2,P_i,P_f) = ieZF(q^2)P^\mu    
\end{equation}
(equation 3.1 in \cite{Davoudiasl:2021mjy}) where $P_i$ and $P_f$ are the initial and final state nucleus momentum, and $F(q^2)$ is the nuclear elastic form factor. Using this to calculate the amplitude will produce a Z$^2$ factor enhancement. For lead, this coherence factor is 6274 producing a total possible contribution to the transverse asymmetry on the level of a few ppm. This could be sufficient to completely cancel out the two-photon exchange contribution to the asymmetry and result in a measurement consistent with zero.

While this argument still requires further theoretical study it highlights the possibility of a BSM connection to this observable. The proponents are in close touch with BSM phenomenologists to firmly establish this link.

\subsection{Proposed experiment}
\label{sec:ProposedExp}
\begin{figure}[!ht]
    \centering
    \includegraphics[width=0.85\textwidth]{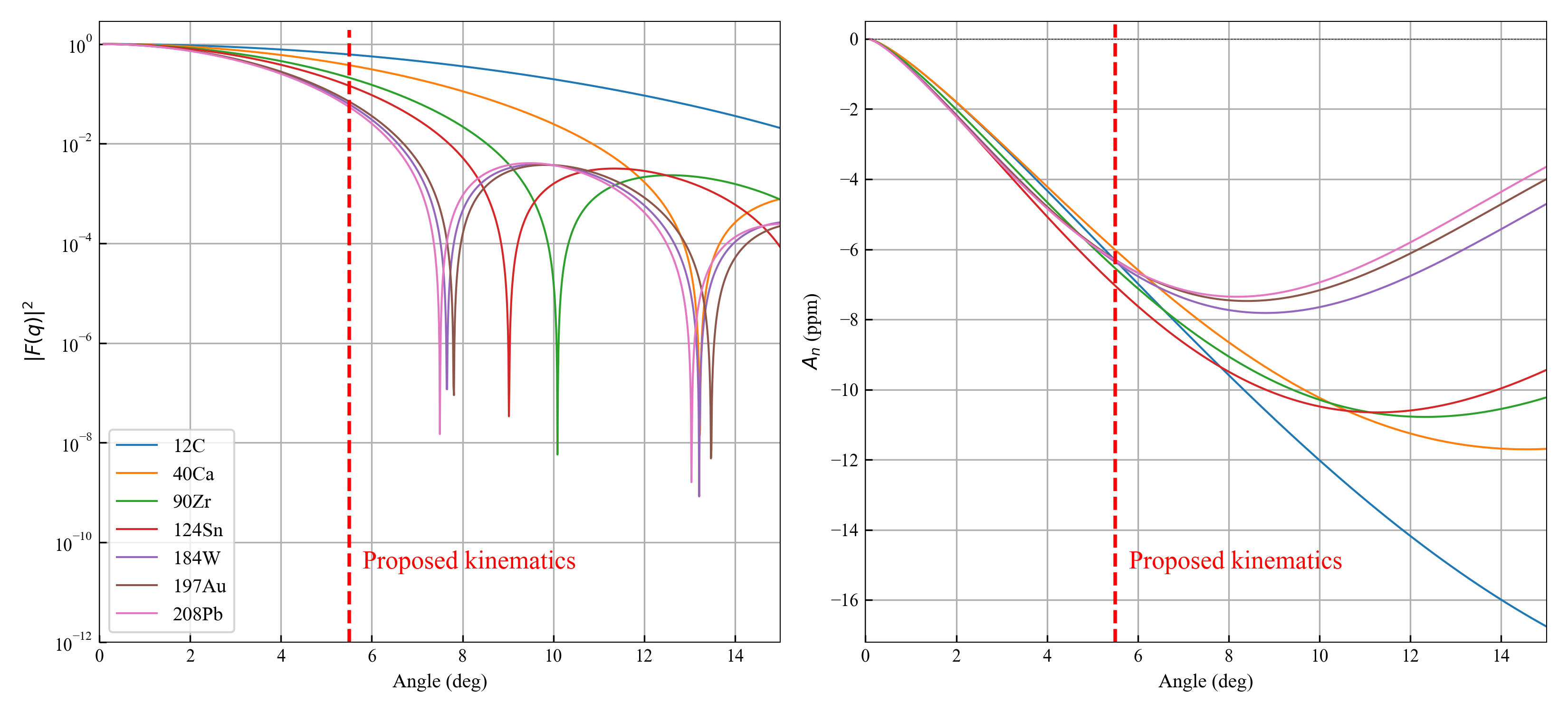} \caption{Form factor square and $A_n$~\cite{mishaPC} for 1 GeV as a function of the scattered angle for the different targets proposed in this experiment. The red dashed line indicates the proposed angle for this experiment.}
    \label{fig:an_new_tgt}
\end{figure}

We propose to measure the nuclear dependence of the beam normal single spin asymmetry for targets with 12 $\leq Z \leq$ 90 at $Q^{2}$ of 0.0092 GeV$^{2}$ at a precision of 0.5 ppm. The proposed targets are summarized in Table~\ref{tab:target_summary}.
Figure~\ref{fig:an_new_tgt} shows the nuclear form factor and $A_{n}$ distributions for 1 GeV as a function of the scattering angle for selected targets. The asymmetries are calculated by Misha Gorchtein based on the approach described in \cite{Koshchii:2021mqq}. The calculation shown in the figure does not include the terms that are not enhanced by the collinear logarithm ($\ln(Q^{2}/m_{e}^2)$), terms that are not suppressed with the extra power of $Q^{2}$, as well as higher order effects in the electromagnetic couplings. 
We note that the proposed targets include nonzero nuclear spin target (such as gold) while the formalism was developed for spin-0 nuclei. The corrections induced for elastic scattering on an unpolarized target with nonzero nuclear spin are of order of the nuclear recoil ($\sim t/M^2$) and are negligible for this experiment~\cite{Koshchii:2021mqq}.

The small angle (5.5 deg) of the proposed kinematics gives a large cross-section while being sufficiently far from the first diffraction minimum. At this kinematics, the asymmetries are expected to be comparable between different targets~(see right panel of Figure~\ref{fig:an_new_tgt}), which makes it ideal for testing the hypotheses described above.

\subsection{Relation to other JLab experiments}
As far as the authors are aware no other approved, conditionally approved, or deferred proposals at JLab address this physics goal.
\section{Experimental setup}
%%Description of standard SHMS spectrometer, detector configuration (tracking detectors, quartz detector + motion system).
This experiment will measure the elastic scattering of transversely polarized electrons off a set of heavy nuclear targets in Hall C. The Super High Momentum Spectrometer (SHMS) will be used to detect the scattered electrons. The SHMS momentum resolution (dP/P) is $2\cdot 10^{-3}$. The angle of the spectrometer will be set to 5.5 deg which is the smallest value at which it can be positioned.

%% Electron beam - energy, current range, polarization, monitoring
The 1.0 GeV electron beam will be polarized transversely. The beam polarization will be measured by a ``spin dance''(see \cite{ahmedZafarSpinDance} for a description of the procedure) using the Moller polarimeter in Hall C. The polarization is expected to be measured to a precision of 1\% or better. The beam current and position will be monitored using the standard Hall C beamline current and position monitors throughout the experiment to constrain and subtract the helicity-correlated asymmetries in the beam. 

% Target
The production targets are listed in Table~\ref{tab:target_summary}. The target selection was dictated by the need to span the Z space between Ca40 and Pb208 while keeping in mind the availability of isotopically pure targets. We note that at $Z\sim 60$ and $Z\sim80$ we selected multiple targets to allow for systematic checks between them~\footnote{The $Z\sim60$ chain of targets all have the same number of neutrons which could be an additional systematic check for models trying to explain the PREX puzzle.}. The targets will be mounted in a copper frame with cryogenic cooling. We note that for some of the targets with a relatively lower melting point (such as tin or lead) we will need to sandwich them between thin graphite or diamond foils similar to what was done for the PREX experiment. 
In addition to the production targets, we will also have a Carbon optics target and a water cell target~\cite{watercell}. The water cell target is used to calibrate the scattering angle using the difference in nuclear recoil between scattering from hydrogen and oxygen. The technique was used in previous parity experiments~\cite{HAPPEX:PhysRevLett.108.102001, PREXCREXAT:2022}.

\begin{figure}[!ht]
    \centering
    \includegraphics[width=0.35\textwidth]{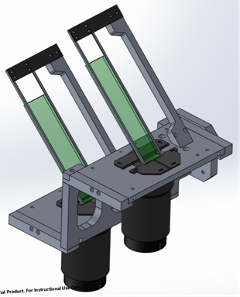}
    \caption{CAD of detector system used in the PREX-2 and CREX experiments. The detectors in this experiment will be mounted in a similar system.}
    \label{fig:Det1}
\end{figure}

% Detector
The detector system would be similar to those that were used in PREX-2 and CREX (see Figure~\ref{fig:Det1} for a CAD model of the PREX-2 and CREX detectors). The scattered electrons will be detected by a pair of radiation-hard fused silica bars (hereafter called quartz). A typical detector will consist of a 200 mm $\times$ 40 mm $\times$ 5 mm quartz bar and an air-coupled PMT. The Cherenkov light produced by the signal electrons inside the detector, as they pass through it, would be converted to an electronic signal by the PMT - leading to an integrating measurement of the electron flux. The detectors will be mounted on a system that can remotely adjust the position and angle of the detector which will allow us to optimize the acceptance of the elastic events. The two detectors will be aligned along the dispersive direction to allow the first one to measure elastic scattering, and the second one a mix of inelastic and elastic events as a cross-check on the systematic backgrounds (see Sec.~\ref{sec:syst} for further discussion).

The existing SHMS detector package will be used to measure the $Q^{2}$ when running the DAQ in counting mode~\cite{HallCPackage} at significantly lower beam currents. To ensure sufficient precision in the reconstruction of kinematic variables we will perform a pointing measurement with a water cell target.

\subsection{Kinematics and rate estimates}
\begin{figure}[!ht]
    \centering
    \includegraphics[width=0.95\textwidth]{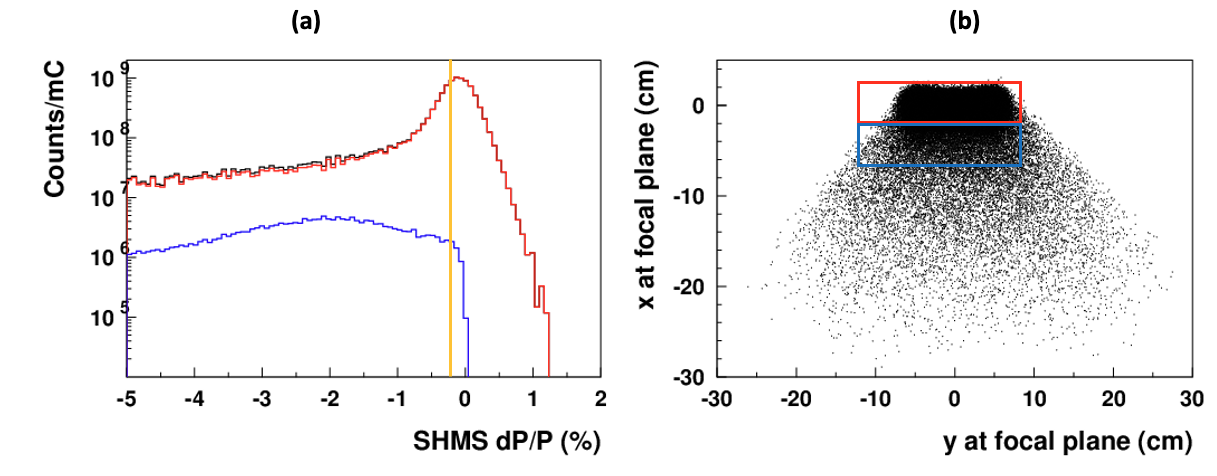}
    \caption{Simulated results for a 3\% radiation length gold target with the SHMS. Panel a: Distribution of SHMS $dP/P$. The red line is for the elastic yield, the blue line is for the quasi-elastic and inelastic yield, and the black line is the total yield. The vertical orange line shows the 0.2\% cut used for rate estimation. Panel b: positional distribution of events in the focal plane. The vertical axis is the dispersive direction. The red and blue rectangles depict a potential arrangement of the two detectors. }
    \label{fig:focalPlane}
\end{figure}

Table~\ref{tab:kin} summarizes the proposed kinematics for this experiment. For the production data, we request 5.1 days of beam time to achieve a statistical precision of 0.5 ppm. The rates are estimated using the Hall C Monte Carlo simulation package (SIMC). The simulation includes the spectrometer acceptance and the effects from multiple scattering, ionization, and radiative energy losses. The detector acceptance is defined to have the signal detector edge at 2 MeV from the momentum peak (equivalent to 0.2\% in Figure~\ref{fig:focalPlane}). We assume an electron beam polarization of 80\%, a current of 30 $\mu$A and the target thickness to be 3\% of a radiation length for all targets. The estimated elastic rates are 200--500 MHz depending on the target and are summarized in Table~\ref{tab:target_summary}.

\begin{table}[!ht]
    \centering
    \begin{tabular}{|l|c|}
    \hline
    $E_{beam}$ (GeV) & 1.0 \\
    $\theta_{lab}$ (deg) & 5.5 \\
    $Q^{2}$ (GeV$^{2}$) & 0.0092 \\
    Beam current ($\mu$A) & 30 \\
    Statistical uncertainty & 0.5 ppm \\
    Systematic uncertainty &  0.2 ppm \\
%    Production data & 4.5 day \\
%    Commissioning & 3.5 days \\
%    \hline
%    Total Beam Time Request & 8 days \\
    \hline
    \end{tabular}
    \caption{Proposed kinematics for this experiment}
    \label{tab:kin}
\end{table}

\begin{table}[!ht]
    \centering
    \begin{tabular}{|c|c|c|c|c|c|c|}
    \hline
    Target & \makecell{Proton\\ number} & \makecell{Thickness\\ ($mg/cm^{2}$)} & \makecell{Beam current\\ ($\mu$A)} & Rate (MHz) & \makecell{Beam time\\ (hours)} & \makecell{Position scan\\ (hours)} \\
    \hline
    $^{12}$C & 6 & 1280.9 & 30 & 494 & 1.76 & 20 \\ 
    $^{40}$Ca & 20 & 483.13 & 30 & 411 & 2.11 & - \\
    $^{90}$Zr & 40 & 301.6 & 30 & 306 & 2.84 & - \\ 
    %$^{107}$Ag & 47 & 266.7 & 30 & &\\ 
    $^{124}$Sn & 50 & 276.1 & 30 & 261 & 3.32 & 20\\ 
    $^{140}$Ce & 58 & 238.3 & 30 & 247 & 3.51 & - \\
    $^{142}$Nd & 60 & 227.5 & 30 & 246 & 3.52 & - \\
    $^{144}$Sm & 62 & 217.7 & 30 & 245 & 3.54 & - \\
    $^{182}$W & 74 & 200.8 & 30 & 216 & 4.02 & - \\ 
    $^{197}$Au & 79 & 193.8 & 30 & 207 & 4.20 & 20\\ 
    $^{208}$Pb & 82 & 191.1 & 30 & 200 & 4.34 & 20\\
    $^{232}$Th & 90 & 182.2 & 30 & 189 & 4.60 & -\\
    \hline
    \multicolumn{6}{|r|}{Total production beam time} & 5.1 days\\
    \hline
    \end{tabular}
    \caption{Production targets and estimated elastic rates}
    \label{tab:target_summary}
\end{table}

\subsection{Elastic peak scans}
Some of the proposed targets (table~\ref{tab:target_summary}) have low-lying inelastic states that could contaminate the elastic $A_n$ measurement. While the inelastic rates will be dwarfed by the elastic rates, their possible\footnote{Predictions for the inelastic state asymmetries were not available in the literature at the time of the submission of this proposal.} large asymmetry contribution could make for a harder interpretation of our results. We propose to collect data with different detector positions to have access to different mixes of inelastic and elastic contributions.

The rates for this experiment allow for the collection of the full statistics for each target in approximately 4 hours (see Table~\ref{tab:target_summary}) by integrating all the scattered electrons in the primary detector. The secondary detector would measure the asymmetry lower in the focal plane with a mix of elastic and inelastic events as a cross-check (see panel b of Figure~\ref{fig:focalPlane}). 
We propose to run two additional measurements where the two detectors would separate the elastic peak with 60\%:40\% and 40\%:60\% (scan along the vertical axis in Figure~\ref{fig:focalPlane} - x at the focal plane). This would allow the primary detector to have a progressively cleaner sample of elastic events. Moreover, the 3 different measurements where the primary detector sees 100\%, 60\%, 40\% of the elastic peak would allow for a deconvolution of the inelastic and elastic asymmetries in the data analysis if one indeed is present. To obtain comparable statistics to the main measurement we estimate that each of these additional runs will require 10h of beam time, totaling 20h of beam time for each target. 

These scans will be performed for four of the production targets: the theoretically well-understood carbon, the medium-Z tin nucleus, and two high-Z nuclei. The first high-Z target would be the lead-208 target for a calibration of the analysis method since it has a large separation between the elastic peak and the first inelastic state. The gold-197 target would be the secondary high-Z target since the first inelastic state sits at only 77~keV away from the elastic peak.

\subsection{Commissioning and auxiliary measurements}
In addition to the production run, we also request 3.5 days of beam time for commissioning and other auxiliary measurements. The additional beam time includes:
\begin{itemize}
    \item Parity quality beam (PQB): The beam will be set up from the injector to the hall with minimal helicity correlated asymmetries. This is a standard procedure that has been established most recently with the PREX-2 and CREX experiments in hall A and previously with the Qweak experiment in hall C. We estimate this would require 8 hours of beam time. Additionally, the beam modulation system in hall C will have to be recommissioned and the detector responses (with different targets) to the beam and energy modulation will need to be established. We estimate this would require an additional 16 hours of commissioning bringing the total for PQB to 24 hours.
    \item Beam polarimetry: As described in the systematic section the polarization of the beam will need to be measured in a longitudinal configuration to get the optimal Wien angle and total beam polarization with the M\o ller polarimeter in the hall. A secondary measurement will need to be made when the vertical polarization is established in the hall to ensure that no longitudinal polarization remains. To reduce systematics we would like to request a high-precision Mott polarimeter measurement in the injector. We estimate this will require 24 hours of beam time.
    \item Spectrometer commissioning: the tune of the spectrometer will be established and the detector positions will be optimized. We estimate this would require 16 hours.
    \item Pointing measurements: We will make use of a water cell target to precisely determine the angle of the spectrometer. We estimate this would require 4 hours of beam time.
    \item $Q^{2}$ measurements: We will take low current runs to determine the kinematics of the spectrometer setup with the production settings. Two such measurements are envisioned, one before production, and one after production to determine the stability of the measurements. We estimate this would require 8 hours of beam time.
\end{itemize}

\section{Sources of systematics}
\label{sec:syst}
%Measuring a 5 ppm asymmetry within 10\% absolute uncertainty requires special attention to controlling systematically. The systematic control should be better than statistical precision (5$\times 10^{-7}$), hence our systematic control goal is $< 10^{-8}$. The potential sources of systematic uncertainties are described in the following subsections. The systematic controls achieved by the recent parity experiments (PREX/CREX) will be used as a reference for this proposal.
The expected TPE contributions for the measurements proposed here (see Figure~\ref{fig:an_new_tgt}) are $\approx$6~ppm. We propose measurements that will reach a precision of 0.5~ppm absolute uncertainty. We expect to reach a systematic control of 0.2~ppm or better. The recent experimental results for PREX-2, CREX, and the $A_n$ measurements associated with these experiments reached a total systematic uncertainty of 0.008~ppm, 0.04~ppm, and 0.09~ppm\footnote{We consider only the systematic obtained for the comparable 1~GeV $A_n$ measurements.} respectively. Comparatively, the systematic goals for these measurements are less stringent. Table~\ref{tab:syst_summary} summarizes the systematic effects that we expect in this experiment.

\begin{table}[!htb]
    \centering
    \begin{tabular}{|l|c|c|}
    \hline
    Effect & Uncertainty [ppb/percent] & PREX-2 $A_n$ achieved [ppb] \\
    \hline
    Inelastic contributions & 130 / 2.2\%& $\leq$10\\
    $A_{\rm false}$ & 100 / 1.6\%  & $\leq$80\\
    Polarization & 70 / 1.1\%& $\leq$60\\
    Detector non-linearity & 50 / 0.8\%& $\leq$30\\
    Target impurities & 50 / 0.8\%& $\leq$40\\
    \hline
    Total & 192 / 3.2\% & $\leq$113\\
    \hline
    Statistical precision & 500 / 8.3\%& - \\
    \hline
    \end{tabular}
    \caption{Summary of target systematic uncertainties. The percent numbers are presented assuming an $A_n$ of 6~ppm.}
    \label{tab:syst_summary}
\end{table}

\paragraph{False asymmetries:} $A_{false}$ can arise from helicity-correlated changes in the beam properties (position, angle, energy) before the interaction. This in turn can create the appearance of an asymmetry in the detector package. We will minimize these effects through the standard techniques used in the recent parity experiments at JLab (minimizing these asymmetries, half-wave flips in the injector) and through the use of the feedback. Finally, we will evaluate the size of the remaining effects by making use of the modulation systems in the Hall C line. 
%We have assigned a systematic of 100~ppb to this correction (higher than the 80~ppb obtained with Pb in the PREX-2 $A_n$ measurement).

\paragraph{Inelastic contributions:} It is known that the nuclear excited states can have a large contribution to $A_n$ measurements. The inelastic states will depend on the target nucleus (see table~\ref{tab:target_inelastic}), however, we plan to minimize such backgrounds by accepting only a fraction of the elastic events. The remaining mixture of elastic and inelastic events will be measured by the secondary detector as a cross-check of our understanding of these states. Thirdly, while some nuclei may have low-lying inelastic states that may still creep into our acceptance, the Qweak Al measurement successfully completed a measurement accepting all the nuclear excited states~\cite{Androic:2020rkw}. The authors of that paper established a procedure to evaluating the rates and possible $A_n$ contributions reaching a total systematic of 2.6\%. Comparatively this experiment will be easier since we will only accept some of the nuclear excited states. Finally,
the result of this experiment does not hinge on any single nucleus measurement but rather on the measurement of the collective trend as a function of $Z^2$. These results could bring additional theoretical interest from the community to better understand the inelastic states for these nuclei.
\begin{table}[!ht]
    \centering
    \begin{tabular}{|c|c|c|c|}
    \hline
    Target & Ground state & Energy of the 1st  & Spin parity of the \\
     & spin parity [$J\pi$] & excited state [MeV]  & 1st excited state [$J\pi$]\\
    \hline
    $^{12}$C & 0+ &4.44& 2+\\ 
    $^{40}$Ca & 0+ &3.35& 0+\\
    $^{90}$Zr & 0+ &1.76& 0+\\ 
    $^{124}$Sn & 0+ &1.13& 2+\\ 
    $^{140}$Ce & 0+ &1.60&2+\\
    $^{142}$Nd & 0+ &1.58&2+\\
    $^{144}$Sm & 0+ &1.66&2+\\
    $^{182}$W & 0+ &0.10&2+\\ 
    $^{197}$Au & 3/2+ &0.08&1/2+\\
    $^{208}$Pb & 0+ &2.61&3-\\
    $^{232}$Th & 0+ &0.05&2+\\
    \hline
    \end{tabular}
    \caption{Production targets and their first nuclear excited states.}
    \label{tab:target_inelastic}
\end{table}

\paragraph{Polarization:} Recent measurements with the M\o ller polarimeter have easily reached 1\% systematic precision. Additionally, we assume a 3 degree precision for the setup of the Wien angle to provide vertical polarization in hall C. The total amount of beam polarization will be measured with the electron beam polarized longitudinally in the hall. Additional measurements will be made to establish the correct Wien angle setting. Once that is complete, the Wien angle will be rotated 90 degrees and another measurement will be done with the M\o ller polarimeter to confirm zero remaining longitudinal polarization. 
%We have assigned a systematic uncertainty of 

\paragraph{Detector non-linearity:} The quartz detector response as a function of electron current can have non-linearities that can impact the measurement. We plan to use similar detectors as were used in the PREX-2 and CREX experiments which reduced these effects significantly.

\paragraph{Target impurities:} Some of the targets (such as lead or tin) will need graphite backing to ensure they will withstand the high current beam. Additionally, they may contain mixtures of other isotopes. This systematic is associated with the subtraction of the asymmetry contamination to these targets. 

\section{Projected results}
\begin{figure}[!ht]
    \centering
    \includegraphics[width=0.9\textwidth]{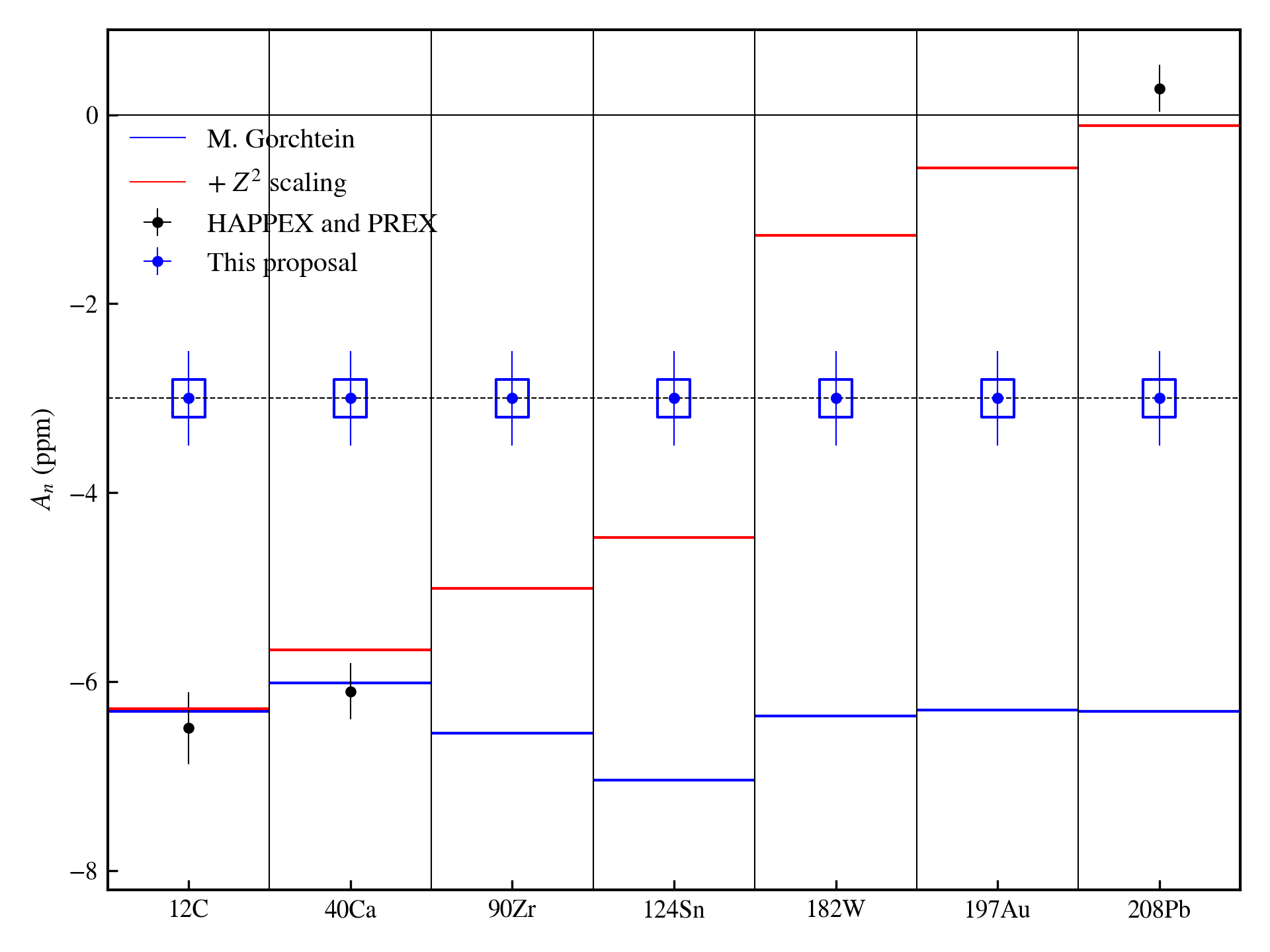}
    \caption{Projected results for $A_{n}$ (blue points) along with calculations and experimental data (black points) for $^{12}$C, $^{40}$Ca, $^{208}$Pb from PREX~\cite{Abrahamyan:2012cg, PREXCREXAT:2022} taken at similar kinematic regions. The data for $^{12}$C and $^{208}$Pb were taken at $Q^{2}$ of 0.00984 GeV$^2$ and 0.00881 GeV$^2$, respectively while $^{40}$Ca data was taken at $Q^{2}$ = 0.0065 GeV$^2$. The blue line represents the calculated asymmetries by M. Gorchtein~\cite{mishaPC} as described in ~\ref{sec:ProposedExp}. The red line is for the calculations including the radiative corrections suggested from \cite{PREXCREXAT:2022}.}
    \label{fig:proj}
\end{figure}

Figure~\ref{fig:proj} shows the projected $A_{n}$ results with 0.5 ppm uncertainty for selected targets together with the existing data from \cite{Abrahamyan:2012cg, PREXCREXAT:2022} and theoretical predictions. As shown in the figure, the theoretical calculations based on two-photon exchange suggest no dependence of the asymmetry on Z at the same time as the calculation is in disagreement with the data for $^{208}$Pb. We also compare the projected results with the calculations assuming a $Z^2$ correction with the following form 
\begin{equation}
    A_{n}\approx A_{0}(Q)(1-C\cdot Z^2\alpha),
\end{equation}
as suggested in \cite{Abrahamyan:2012cg}. $A_0 (Q)$ is the TPE theoretical prediction (the blue lines in the figure), and $C\approx 0.2$ is an empirical constant that was determined based on the existing data from Jefferson Lab and Mainz. One can clearly see that the importance of the heavy nuclei with $Z \geq 40$ for testing the hypotheses suggested in Sec.~\ref{sec:hypotheses}.

\section{Summary}
We propose the measurements of beam normal single spin asymmetries in elastic scattering of the transversely polarized electron from target nuclei with 12 $\leq Z \leq$ 90 to study the nuclear dependence of the asymmetry. In particular, this experiment will allow us to investigate if the proposed $Z^{2}$ scaling will hold, but more importantly provide data to advance the understanding of these asymmetries. The experiment will be carried out in Hall C using the SHMS at $Q^{2}$ = 0.0092 (GeV$^2$).

We request total of 8.6 PAC days of transversely polarized electron beam at 1.0 GeV. This includes 5.1 day of production data taking to achieve a precision of 0.5 ppm and 3.5 days of commissioning and auxiliary measurements.

\begin{table}[!ht]
    \centering
    \begin{tabular}{|l|c|}
    \hline
    Production & 5.1 days \\
    Commissioning & 2.5 days\\ 
    Auxiliary measurements & 1 day\\
    \hline
    Total Beam Time Request & 8.6 days\\
    \hline
    \end{tabular}
    \caption{Requested beam time for this experiment}
    \label{tab:BeamTime}
\end{table}

\newpage
\bibliographystyle{unsrt}
\bibliography{bib.bib}

\end{document}